\documentclass[prl,final,twocolumn,showpacs,showkeys]{revtex4}
\usepackage{amsmath}
\usepackage{graphicx}
\usepackage{amsfonts}
\usepackage{amssymb}
\usepackage{subfigure}
\begin{document}

\newcommand{\subs}[1]
{ 
	\mbox{\scriptsize{#1}}
}

\newcommand{\neswarrow}{\mathrel{\text{\tiny{$\nearrow$\llap{$\swarrow$}}}}}
\newcommand{\nwsearrow}{\mathrel{\text{\tiny{$\nwarrow$\llap{$\searrow$}}}}}

\title{All-Resonant Control of Superconducting Resonators}

\author{Frederick W. Strauch} \email[Electronic address: ]{Frederick.W.Strauch@williams.edu}
\affiliation{Williams College, Williamstown, MA 01267, USA}
\date{\today}

\begin{abstract}
An all-resonant method is proposed to control the quantum state of superconducting resonators.  This approach uses a tunable artificial atom linearly coupled to resonators, and allows for efficient routes to Fock state synthesis, qudit logic operations, and synthesis of NOON states.  This resonant approach is theoretically analyzed, and found to perform signficantly better than existing proposals using the same technology.   
\end{abstract}
\maketitle
%%%%End of Front Matter%%%%%%%%%%%%%%%%%%%%%%%%%%%%%%%%%%%%%%%%%%%%%%%%%
%%%%%%%%%%%%%%%%%%%%%%%%%%%%%%%%%%%%%%%%%%%%%%%%%%%%%%%%%%%%%%%%%%%%%%%%

Achieving complete control of the quantum state of light is a primary goal in the field of quantum optics and quantum information \cite{HarocheBook}.  The preparation and subsequent interaction of individual photons for quantum communication and computation in the optical domain remains a challenging enterprise \cite{Kok07}.  By contrast, excitations of the electromagnetic modes of superconducting coplanar waveguide resonators can be readily prepared and manipulated using Josephson junction circuits \cite{You11}.  These microwave photons have recently been proposed as key ingredients in a superconducting quantum computer \cite{Mariantoni11}.  Finding the fastest and most efficient way to control these modes is an outstanding problem.

Great progress has been made by using a tunable artificial atom, the superconducting phase qubit, to excite and transfer excitations to and from superconducting resonators \cite{Sillanpaa2007}.  Subsequent experiments have prepared individual Fock states \cite{Hofheinz2008}, their superpositions \cite{Hofheinz2009}, and entangled states of two such resonators \cite{Wang11}.  Despite this significant progress, complete control of these resonators \cite{Strauch10,Strauch11} appears to require a Fock-state-selective interaction.   Such number-state-dependent interactions were first seen by Schuster {\it{et al.}} \cite{Schuster07} and exploited for Fock-state measurement \cite{Johnson10} in transmon qubit devices, while Fock states have also prepared using sideband transitions \cite{Leek2010}.  These latter experiments used dispersive (off-resonant) coupling of the qubit to a resonator,  while the former experiments utilized a carefully chosen sequence of qubit operations performed off resonance and resonant qubit-resonator swaps.  As resonant interactions are often faster than their off-resonant counterparts, an important question is whether complete control can be achieved using resonant interactions alone.  

In this Letter, I present precisely such an all-resonant method appropriate for superconducting resonators.  This method is shown to be applicable to the synthesis of arbitrary superpositions of Fock states of one and two resonators.  By performing all of the important steps with the qubit on resonance with the resonator, these applications are found to be significantly faster, more efficient, and of higher fidelity than previous proposals.  While this method is designed for superconducting experiments with existing technology, I will present the basic principles using the broadly applicable Jaynes-Cummings Hamiltonian, so that any tunable atom can be used to quickly control any harmonic oscillator mode to which it is coupled.

%%%%%%%%%%%%%%%%%%%%%%%%%%%%%%%%%%%%%%%%%
\begin{figure}[t]
\begin{center}
\includegraphics[width=3in]{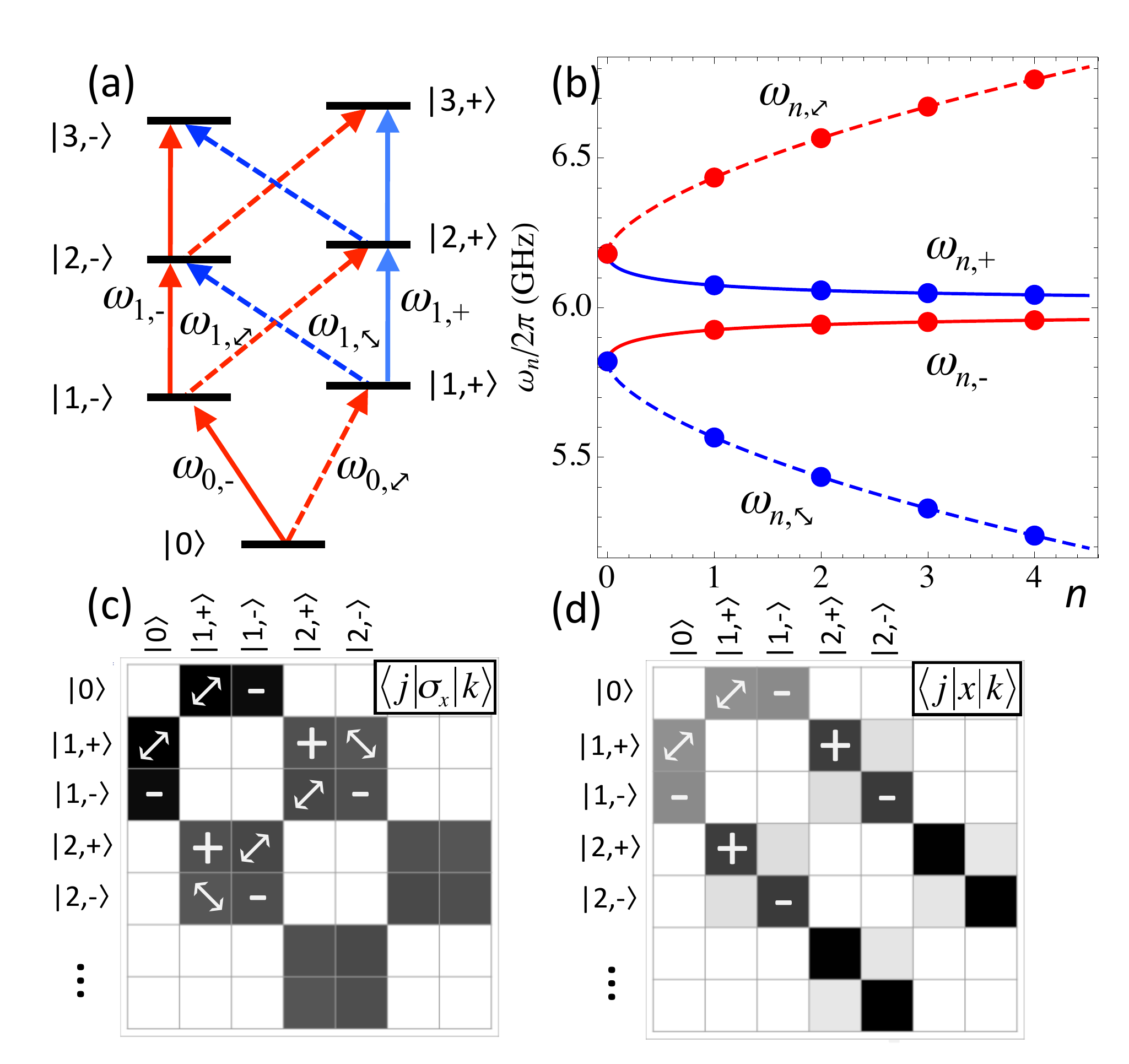}
\caption{(Color online) Jaynes-Cummings Hamiltonian transitions. (a) Energy level diagram, with the four basic transitions $\omega_{n,+}$ (red solid arrows, on left), $\omega_{n,-}$ (blue solid arrows, on right), $\omega_{n,\nwsearrow}$ (blue dashed arrows, right to left), and $\omega_{n,\neswarrow}$ (red dashed arrows, left to right).  (b) Transitions frequencies $\omega_n/2\pi$ as a function of $n$, for $\omega_r/2\pi = 6 \ \mbox{GHz}$, $g/2\pi = 180 \ \mbox{MHz}$, and $\Delta = 0$. (c) Transition matrix elements $\langle j | \sigma_x |k\rangle$ due to a drive on the qubit.  (d) Transition matrix elements $\langle j | x |k\rangle$, with $x = a+a^{\dagger}$, due to a drive on the resonator.  Dark values indicate larger matrix elements, while the corresponding transitions are indicated by the symbols.  }
\label{JC_Diagram}
\end{center}
\end{figure}
%%%%%%%%%%%%%%%%%%%%%%%%%%%%%%%%%%%%%%%%%

{{\it Model}}. The Jaynes-Cummings Hamiltonian \cite{HarocheBook} is
\begin{equation}
\mathcal{H}/\hbar = \omega_q \sigma_+ \sigma_- + \omega_r a^{\dagger}a + g \left(a \sigma_+ + a^{\dagger} \sigma_- \right),
\end{equation}
where $\omega_q$ and $\omega_r$ are the transitions frequencies and $\sigma_+ = |1\rangle \langle 0|$ and $a^{\dagger}$ are the creation operators for the qubit and resonator, respectively, and $g$ is the coupling between the qubit and oscillator.  Famously, this can be diagonalized by the eigenstates
\begin{eqnarray}
|n,-\rangle = \cos \theta_n |0,n\rangle - \sin \theta_n |1,n-1\rangle \nonumber \\
|n,+\rangle = \sin \theta_n |0,n\rangle + \cos \theta_n |1,n-1\rangle
\end{eqnarray}
where $\tan (2\theta_n) = 2 g \sqrt{n}/\Delta$, $\Delta = \omega_q - \omega_r$, and the $|q,n\rangle$ represents a state with the qubit $q = 0 \ \mbox{or} \ 1$ and oscillator number $n$.  The corresponding eigenvalues are given by
\begin{equation}
E_{n,\pm}/\hbar = n \omega_r + \frac{1}{2} \left( \Delta \pm \sqrt{ \Delta^2 + 4 n g^2} \right).
\end{equation}
Note that the ground state has $E_0 = 0$ and is given by $|0\rangle = |0,0\rangle$.  The energy level diagram is shown in Fig. 1(a), along with several transitions, whose frequencies are given by
\begin{eqnarray}
\omega_{n,\pm} &=& \omega_r \pm \frac{1}{2} \left( \sqrt{\Delta^2 + 4 (n+1) g^2} - \sqrt{\Delta^2 + 4 n g^2} \right), \nonumber \\
\omega_{n,\neswarrow} &=& \omega_r + \frac{1}{2} \left(\sqrt{\Delta^2 + 4 (n+1) g^2} + \sqrt{\Delta^2 + 4 n g^2} \right), \nonumber \\
\omega_{n,\nwsearrow} &=& \omega_r - \frac{1}{2} \left(\sqrt{\Delta^2 + 4 (n+1) g^2} + \sqrt{\Delta^2 + 4 n g^2} \right). \nonumber \\
\end{eqnarray}
Roughly speaking, the transitions $\omega_{n,\pm}$ correspond to excitations of the resonator, while $\omega_{n,\neswarrow}$ corresponds to rotations of the qubit.  This is most clearly seen in the dispersive regime $\Delta \gg g$, where $\omega_{n,\pm}~\approx~\omega_r~\pm~\left[g^2/\Delta - (2n+1) g^4/\Delta^3 \right]$ is the Kerr-shifted resonator transition and $\omega_{n,\neswarrow} \approx \omega_q + (2 n +1) g^2/\Delta$ is the Stark-shifted qubit transition.  However, in the resonant regime, when $\Delta = 0$, the latter transition exhibits a significantly stronger dependence on $n$, as shown in Fig. 1 (b), while a new transition $\omega_{n,\nwsearrow}$ becomes possible by driving the qubit by a term $\mathcal{H}_{\subs{drive}} = \hbar f(t) \sigma_x$.  This can be seen by the matrix elements of $\sigma_x$ and $x = a + a^{\dagger}$, as shown in Fig. 1(c) and (d).   

Previous transmon experiments have probed only a subset of the allowed transitions in Fig. \ref{JC_Diagram}.  The dispersive $\omega_{n,\neswarrow}$ transitions were studied in \cite{Schuster07}, while the resonant $\omega_{1,\pm}$ transitions were studied in \cite{Fink2008}.  Observing the resonant diagonal transitions $\omega_{n,\nwsearrow}$, a novel type of sideband transition, would be a further test of the Jaynes-Cummings model for superconducting circuits, and offers a new path toward Fock state synthesis and control.

Previous theoretical studies have considered sudden shifts of the qubit frequency from the dispersive regime to the resonant regime to swap excitations from the qubit to the resonator $|1,n\rangle \to |0,n+1\rangle$, and with qubit transitions performed in the dispersive regime.   Here I consider controlling the system when the qubit and resonator are resonant, by driving the drive the $\omega_{n,\neswarrow}$ and $\omega_{n,\nwsearrow}$ transitions simultaneously.  As will be shown below, these transitions allow for fast, high-fidelity control of the state.  Of course, for $\Delta = 0$, the eigenstates of the qubit-resonator system are entangled.  To decouple the resonator, one could adiabatically shift the qubit frequency to the dispersive regime.  However, for many applications, this procedure is unnecessary, so I will focus on controlling the system in the resonant regime.

{\it{Fock state preparation}}.  The first task to consider is the preparation of a Fock state $|0,N\rangle$, which by using the adiabatic decoupling reduces to the transformation $|0\rangle \to |N,-\rangle$.  The simplest means to do so would involve $N$ steps, each performing a two-level transition by alternately driving the transitions $\omega_{n,\neswarrow}$ and $\omega_{n+1,\nwsearrow}$, whose precise sequence depends on whether $N$ is even or odd.  However, since these transitions are sufficiently separate in frequency, these transitions can be driven simultaneously.  By choosing an appropriate set of amplitudes, one can perform this transformation in a {\it single} step.

Specifically, I consider driving functions of the form
\begin{equation}
f(t) = \sum_{n=1}^N \left[ A_{n}(t) \cos (\omega_n t) + B_n(t) \sin (\omega_n t) \right],
\end{equation}
where the slowly-varying envelope functions $A_n(t)$ and $B_n(t)$ and frequencies $\omega_n$ are chosen to optimize the transition.  An analytical solution to this problem can be found in the rotating wave approximation, by letting $A_n(t) = \Omega_n$ and $B_n(t) = 0$, where the set of amplitudes are given by
\begin{equation}
\Omega_{n} = \Omega_0 \left\{ \begin{array}{cl}
\sqrt{2 N} & \mbox{for} \ n=1, \\
 2 \sqrt{n (N+1-n)} & \mbox{for} \ 1< n \le N,
\end{array} \right.
\end{equation}
and the frequencies are given by 
\begin{equation}
\{ \omega_n \} = \left\{ \begin{array}{cl} \omega_{0,-}, \omega_{1, \neswarrow}, \cdots, \omega_{N-1, \nwsearrow} & \mbox{for} \ N \ \mbox{odd}, \\
\omega_{0,\neswarrow}, \omega_{1, \nwsearrow}, \cdots, \omega_{N-1, \nwsearrow} & \mbox{for} \ N \ \mbox{even}.
\end{array} \right.
\end{equation}
This choice of frequencies follows a zig-zag pattern up the Jaynes-Cummings ladder, while the amplitudes are chosen so that, in a rotating frame \cite{Footnote}, the driving Hamiltonian is equivalent to an angular momentum operator for a spin-$N/2$ system.  This has the effect of driving a perfect rotation from $|0\rangle \to |N,-\rangle$ in a time $T = \pi / \Omega_0$.  This analytical solution to a population transfer problem was first studied by Cook and Shore \cite{Shore79}, and was exploited in the phase qudit experiment \cite{Neeley2009}; applications to perfect state transfer in qubit networks have also been studied \cite{Christandl04}.

How does this single-step transition (with simultaneous driving) compare with a multi-step transition (with sequential driving)?   To meaningfully answer this question, we must limit the drive amplitudes $\Omega_n$ to common value of $\Omega_{\subs{max}}$.  The sequential, multi-step transitions would then take a time of $T_{\subs{multi}} = (\sqrt{2} + 2 (N-1)) \pi / \Omega_{\subs{max}} \approx 2 N \pi/\Omega_{\subs{max}}$ (the factors of $\sqrt{2}$ and $2$ are due to the transition matrix elements).  For the single-step transition, Eq. (6) shows that $\Omega_n < \Omega_0 (N+1)$, so that $T_{\subs{single}} \approx N \pi / \Omega_{\subs{max}}$, a factor of two better than the multi-step transition.  Studies of perfect state transfer \cite{Yung06} show that this is in fact an optimal solution, given constant and bounded amplitudes.

\begin{figure}
\begin{center}
\includegraphics[width=3.5in]{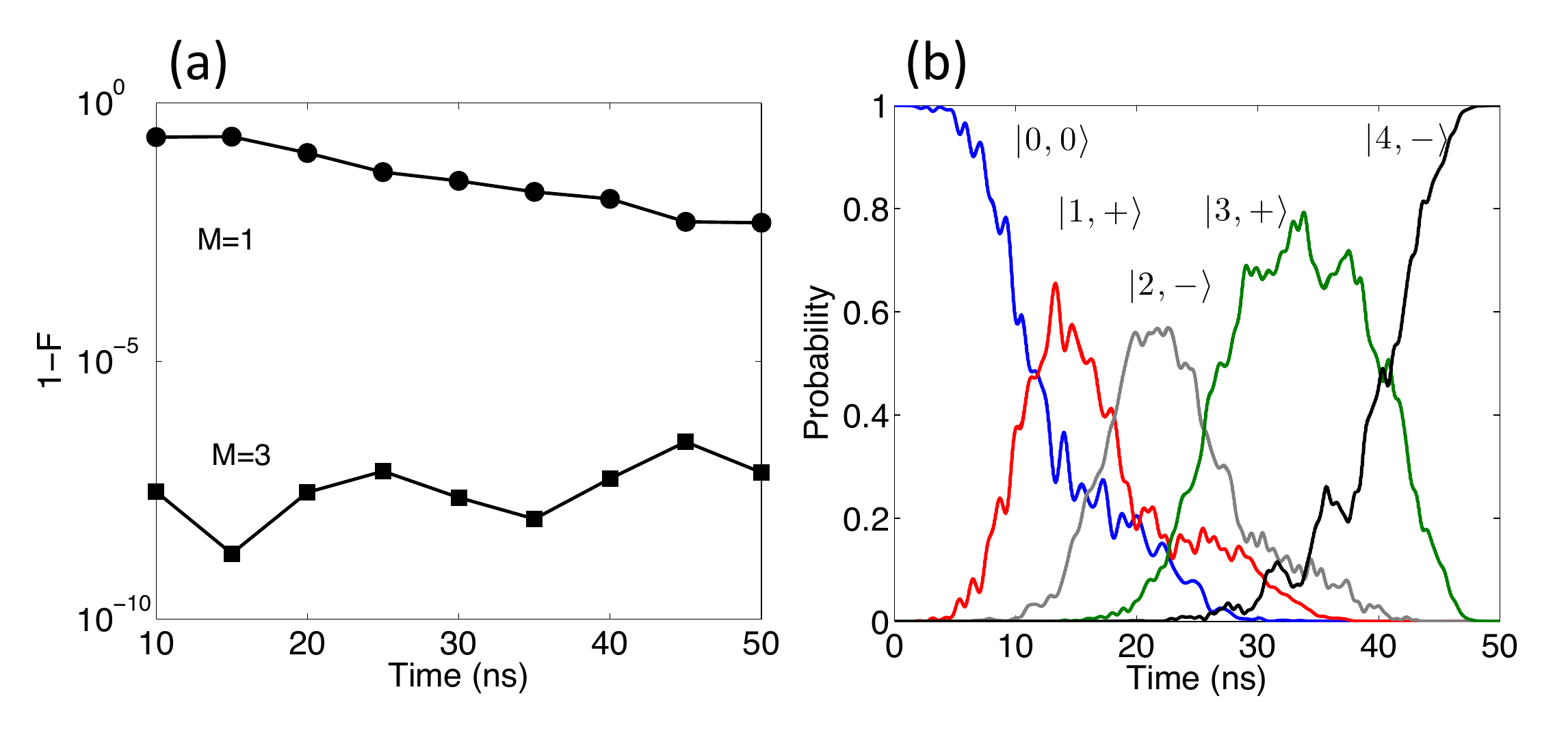}
\caption{(a) Error as a function of the pulse time $T$ for the Fock state transitions $|0\rangle \to |N,-\rangle$, for $N=4$ and using the same parameters as Fig. 1.  The circles are for a numerically optimized two-quadrature pulses with $M=1$ Fourier component, while the squares are for $M=3$ Fourier components. (b) Time-dependent probabilities $|\langle v | \Psi(t)\rangle|^2$ as a function of time $t$ for an $M=3$ pulse with $T=50 \ \mbox{ns}$, where $|v\rangle = |0,0\rangle, |1,+\rangle, |2,-\rangle, |3,+\rangle, \ \mbox{and} \ |4,-\rangle$.  }
\label{Fock_Preparation}
\end{center}
\end{figure}

For the system at hand, of course, there are more than $N$ transitions involved, opening up a signficant source of error.  Isolating a single transition in a multlevel system has inspired a number of control methods \cite{Tian2000,Motzoi09,Forney10}.  I have extended the two-quadrature approach of \cite{Motzoi09} to the multilevel transition case studied here by numerically simulating the time-dependent Schr{\"o}dinger equation $i \hbar \partial_t |\Psi (t)\rangle = (\mathcal{H}+ \mathcal{H}_{\subs{drive}}(t) |\Psi(t)\rangle$, $|\Psi(0)\rangle = |0\rangle$ with the following envelope functions
\begin{eqnarray}
A_n(t) &=& \sum_{k=1}^M a_{n}^{(k)} \left[ 1 - \cos (2 \pi k t/T) \right], \nonumber \\
B_n(t) &=& \sum_{k=1}^M b_{n}^{(k)} \sin( 2 \pi k t /T),
\end{eqnarray}
for various values of the total time $T$ and number of Fourier components $M$.  For each such case, I numerically optimized the fidelity $\mathcal{F} = |\langle N,-|\Psi(T)\rangle|^2$ to find optimal Fourier components $a_{n}^{(k)}, b_{n}^{(k)}$ and driving frequencies $\omega_n$ \cite{Footnote}.   Figure \ref{Fock_Preparation}(a) shows the resulting error $1-\mathcal{F}$ as a function of $T$ of such an optimization performed for $N=4$ and for $M = 1 \ \mbox{and} \ 3$.  The time-dependent probabilities for an optimized transition with $T=50 \ \mbox{ns}$ and $M=3$ Fourier components are shown in Fig. \ref{Fock_Preparation}(b); the optimized frequencies and Fourier components can be found in the Supplemental Information.   For $M=1$, the $T=50 \ \mbox{ns}$ pulse has $\mathcal{F} > 0.99$, while for $M=3$ extraordinarily high fidelity transitions are possible in as little as $10 \ \mbox{ns}$.  This result highlights the power of this all-resonant approach, being significantly faster than the $120 \ \mbox{ns}$ taken by the Rabi-swap sequence of \cite{Hofheinz2008} (which used $\Omega_0/2\pi \approx 25 \ \mbox{MHz}$ and $g/2\pi \approx 18 \ \mbox{MHz}$).   

%%%%%%%%%%%%%%%%%%%%%%%%%%%%%%%%%%%%%%%%%
%%%%%%%%%%%%%%%%%%%%%%%%%%%%%%%%%%%%%%%%%

This method has several advantages over alternative proposals for resonator control.  First, there is no need to shift the qubit frequency to exchange quanta between the qubit and the resonator; the $\omega_{n,\nwsearrow}$ transitions accomplish this directly, much like sideband transitions in ion-trap systems.  This allows for larger qubit-resonator couplings.  Second, this approach can also be pursued for large couplings and photon numbers, whereas dispersive manipulations are limited to $N < n_{\subs{crit}} = 4 \Delta^2 / g^2$.  While the multilevel structure of superconducting qubits will also limit the transitions  \cite{Tian2000,Motzoi09,Forney10}, this effect is minimized when $g < |\omega_{01} - \omega_{12}|$, precisely when the two-level qubit approximation is appropriate for the Jaynes-Cummings Hamiltonian.  Finally, the large frequency separation in the resonant regime allows for much faster transitions than using dispersive transitions (on the qubit or resonator).  That is, to avoid non-resonant transitions one should require $\Omega_{\subs{max}} \le \Delta \omega$, where $\Delta \omega$ is the smallest frequency separation in the problem.  Using the expressions of Eq. (4), the frequency separations in the resonant regime are $\Delta \omega_{\neswarrow, \nwsearrow} \sim g/n^{1/2}$,  $\Delta \omega_{\pm} \sim g/n^{3/2}$, much larger than those in the dispersive regime $\Delta \omega_{\neswarrow} \sim 2 g^2/\Delta$ and $\Delta \omega_{\pm} \sim 2 g^4/\Delta$ (since $g \ll \Delta$).  For the case of Fig. \ref{Fock_Preparation}(b), the Fourier components satisfy $|a_{n}^{(k)}|, |b_{n}^{(k)}| < \Delta \omega/2\pi \approx 237 \ \mbox{MHz}$ for pulse times $T > 10 \ \mbox{ns}$ \cite{Footnote}.  By using the transitions with the largest frequency separation, the control pulse can be performed as quickly as possible.

This time advantage is particularly important in the presence of decoherence.  Following the analysis of \cite{Strauch12}, and letting the qubit and resonator have dissipation times of  $T_q$ and $ T_r$, respectively, the fidelity of this transition is $\mathcal{F} \approx e^{ -T/(T _Tq)} e^{- N T / (2 T_r)}$ for both the single-step procedure and the procedure of \cite{Hofheinz2008}.  By reducing the total time of the sequence by a factor of two (or more), the procedure described above will achieve higher fidelity.     

\begin{figure}
\begin{center}
\includegraphics[width=3in]{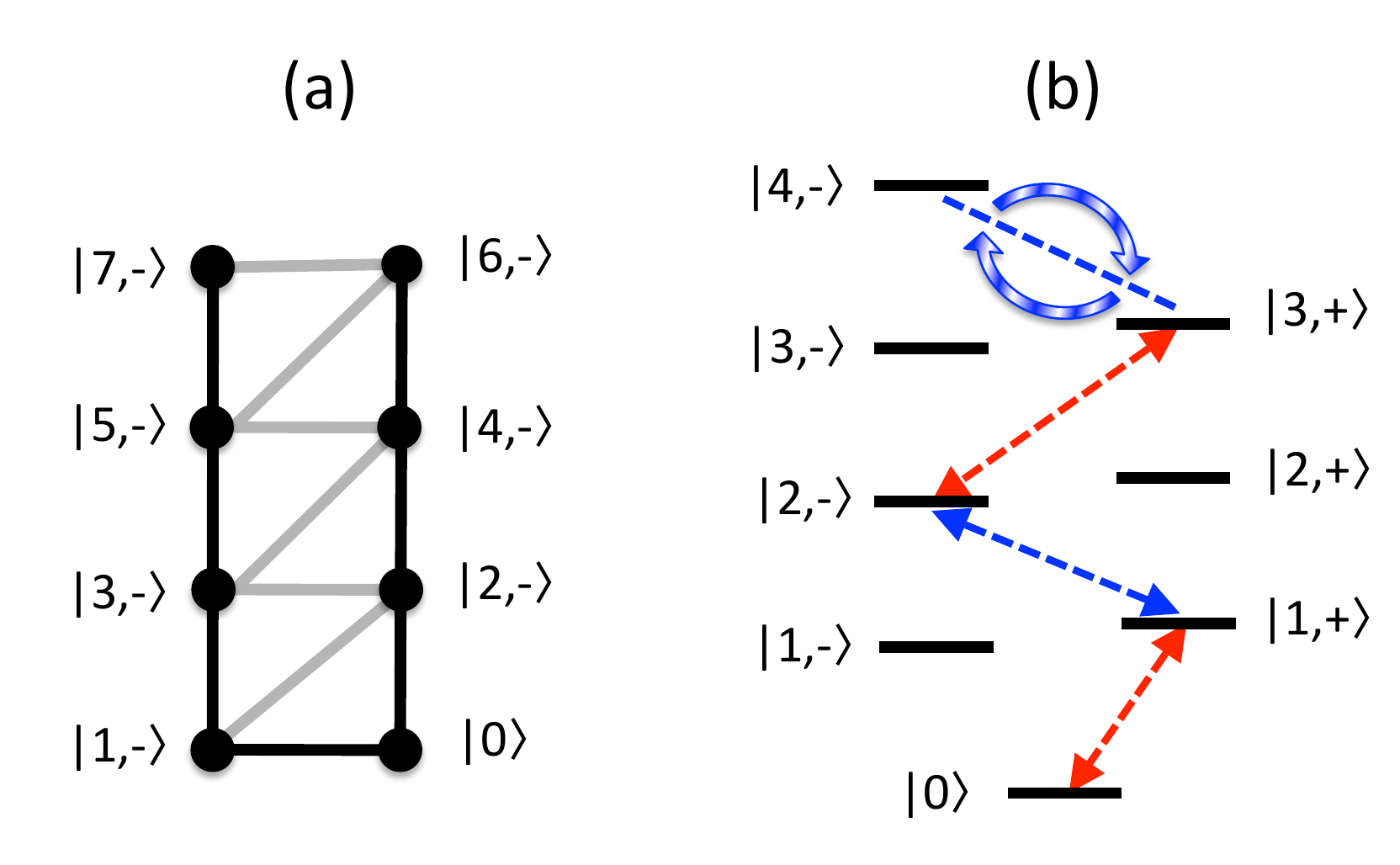}
\caption{(Color online) Qudit operations with the Jaynes-Cummings ladder.  (a) Coupling graph for the states $|0\rangle$ and $|n,-\rangle$, for $n=1 \to 7$. The darker edges correspond to couplings through the transitions $\omega_{\neswarrow, \nwsearrow}$, while the lighter edges correspond to the transitions $\omega_{\pm}$.  The intermediate states $|n,+\rangle$ have been omitted for simplicity.   (b) Schematic rotation $\mathcal{R}_{0,4}(\theta)$, consisting of the sequence $\mathcal{R}_{0,3}(\pi) \mathcal{R}_{3,4}(\theta) \mathcal{R}_{0,3}(\pi)$, whose second step is indicated by the curved arrows (see text). }
\label{Qudit_op}
\end{center}
\end{figure}

{\it Qudit operations}.  The Fock states of the resonator can be used as an effective $d$-level system, called a qudit \cite{Strauch11}.  Arbitrary operations on the qudit can be composed by rotations $\mathcal{R}_{j,k}(\theta)$:
\begin{equation}
\mathcal{R}_{j,k}(\theta) = \exp \left[ -i \frac{\theta}{2} \left( |j\rangle \langle k| + |k\rangle \langle j| \right) \right].
\end{equation}
The criterion for such arbitrary control is a connected coupling graph of such rotations \cite{Brennen05}.  The coupling graph for the Jaynes-Cummings ladder is shown in Fig. \ref{Qudit_op}(a), which clearly satisfies this criterion.  Here all operations can be performed using the transitions $\omega_{\neswarrow, \nwsearrow}$, which are indicated by the dark edges; the intermediate states $|n,+\rangle$ have been omitted for simplicity.  As there are many possibilities for simultaneous transitions, there are many opportunities for qudit logic synthesis.  

%%%%%%%%%%%%%%%%%%%%%%%%%%%%%%%%%%%%%%%%%
%%%%%%%%%%%%%%%%%%%%%%%%%%%%%%%%%%%%%%%%%

As a simple example, consider the rotation $\mathcal{R}_{0,4}(\theta)$.  This is not directly possible, as indicated by the lack of an edge between $|0\rangle$ and $|4,-\rangle$ in the coupling graph, but this rotation can nevertheless be performed by the sequence $\mathcal{R}_{0,3}(\pi) \mathcal{R}_{3,4}(\theta) \mathcal{R}_{0,3}(\pi)$, as illustrated in Fig. \ref{Qudit_op}(b).  The first and final rotations use the simultaneous control pulses described above, which implement the swap $|0\rangle \leftrightarrow |3,-\rangle$, while state $|2,-\rangle$ remains unaffected.  This sequence can be generalized to implement any rotation $\mathcal{R}_{j,k}(\theta)$ in three steps.  This represents a significant advance over the scheme presented in \cite{Strauch11}, which would require nineteen steps (with four in the dispersive regime).  

{\it NOON state synthesis}.  As a final example, I consider the generation of entangled states between two superconducting resonators, of the form
\begin{equation}
|\Psi_{\subs{NOON}}\rangle = \frac{1}{\sqrt{2}} \left( |N,0\rangle + |0,N\rangle \right).
\end{equation}
Methods to synthesize such ``NOON'' states have been proposed in \cite{Strauch10} and \cite{Merkel10}, the latter implemented in \cite{Wang11}.  Each of these methods use a sequence of $N$ qubit rotations and swaps.  I now show how the approach of \cite{Merkel10,Wang11} can be simplified by the single-step Fock procedure.

This method begins by preparing two qubits in the entangled state
\begin{equation}
|\Psi_0\rangle = \frac{1}{\sqrt{2}} \left( |1\rangle_A |0\rangle_B + |0\rangle_A |1\rangle_B \right),
\end{equation}
with each qubit coupled to a resonator ($A$ or $B$) in its ground state.  By adiabatically moving qubit $A$ ($B$) onto resonance with resonator $A$ ($B$), the joint state is mapped to
\begin{equation}
|\Psi_1\rangle = \frac{1}{\sqrt{2}} \left( |1,+\rangle_A |0\rangle_B + |0\rangle_A |1,+\rangle_B \right).
\end{equation}
For $N$ even, one can simply drive the transition $|1,+\rangle \to |N,-\rangle$ on $A$ and $B$ in parallel, and move the qubits off resonance, yielding $|\Psi_{\subs{NOON}}\rangle$ in a single step.   For $N$ odd, one must first drive the transition $|1,+\rangle \to |1,-\rangle$ (which leaves $|0\rangle$ unchanged), and then drive the transition $|1,-\rangle \to |N,-\rangle$, yielding $|\Psi_{\subs{NOON}}\rangle$ in two steps.  

By using the all-resonant transitions, this method outperforms previous proposals in two other ways.  First, this method can be performed much faster than the first proposal \cite{Strauch10}, which used the dispersive number-state-dependent transitions.  Second, this method uses a two-level system to control the Fock states, as opposed to using a three-level system to selectively excite the resonators \cite{Merkel10,Wang11}.  As shown in \cite{Strauch12}, the use of higher excited states leads to a lower fidelity.  Thus, this approach is faster, achieves a higher fidelity, and uses fewer steps than previous proposals. 

{\it{Conclusion}}.  There are many applications of entangled resonators for quantum measurement, Bell inequality tests, and quantum information, which have been described elsewhere \cite{Strauch12}.  The all-resonant approach presented here provides an attractive route to enabling such applications, using existing technology.  

There are two limitations to this approach, to which we now turn.  To decouple the qubit-resonator state, we might use an adiabatic mapping of the uncoupled qubit-resonator state $|0,n\rangle$ to the Jaynes-Cummings states $|n,-\rangle$.  One means to alleviate this is to use a three-level system such that $\omega_r = \omega_{12}$ \cite{Wang11,Strauch11} and apply resonant pulses to perform the mapping $|0,n\rangle \to |\tilde{n},-\rangle = (|1,n\rangle - |2,n-1\rangle)/\sqrt{2}$, followed by a control sequence on the states $|\tilde{n},-\rangle$.  An alternative is to use the Kerr effect alone, as recently analyzed in \cite{Nigg12}, and drive the $\omega_{n,-}$ transitions, albeit at a slower rate.  The second limitation is the direct excitation of qubit states, which thwarts taking full advantange of superconducting resonators, namely their potential for longer coherence times.  However, current experiments have shown that qubits can have coherence times approaching $100 \ \mu \mbox{s}$ \cite{Paik2011}, so that the method proposed here should allow for complex control of states with photon numbers of ten or more.  

In conclusion, I have proposed an all-resonant scheme to perform a range of quantum control protocols for superconducting resonators.  This approach has many advantages over previous studies, and extensions may unlock the power of on-chip microwave photons.  Finally, many of these ideas can be directly applied to other systems of light interacting with real and artificial atomic systems.

\acknowledgments
I gratefully acknowledge discussions with K. Jacobs and R. W. Simmonds.  This work was supported by the NSF under Project No. PHY-1005571.

%\bibliography{report}

\newpage

\begin{widetext}

\section{Supplemental Material}

\subsection{Introduction}

In this supplement we detail the multi-frequency control pulses used in the text.  The eigenstates of the resonant Jaynes-Cummings Hamiltonian (with $\hbar = 1$) 
\begin{equation}
\mathcal{H}_0 = \omega \sigma_+ \sigma_- + \omega a^{\dagger}a + g \left(a \sigma_+ + a^{\dagger} \sigma_- \right),
\end{equation}
are $|0,0\rangle$ and 
\begin{equation}
|n,\pm\rangle = \frac{1}{\sqrt{2}} \left( |0,n\rangle \pm  |1,n-1\rangle \right).
\end{equation}
where $|q,n\rangle = |q\rangle_{\mbox{\scriptsize{qubit}}} \otimes |n\rangle_{\mbox{\scriptsize{oscillator}}}$.  For $n>0$, the eigenstates have the nonzero matrix elements
\begin{equation}
\langle n+1, + | \sigma_x |n, - \rangle = - \langle n+1, - | \sigma_x |n, + \rangle = \frac{1}{2},
\end{equation}
while the $n=0$ state has the matrix elements
\begin{equation}
\langle 1, \pm | \sigma_x |0,0\rangle = \pm \frac{1}{\sqrt{2}}.
\end{equation}

To use the spin rotation method to implement the Fock state preparation $|0\rangle \to |N,-\rangle$ described in the text, we first define the states
\begin{equation}
\{ |v_n\rangle \} = \left\{ \begin{array}{ll}
|0\rangle, |1,-\rangle, |2,+\rangle, \cdots, |N,-\rangle & \mbox{for} \ N \ \mbox{odd}, \\
|0\rangle, |1,+\rangle, |2,-\rangle,  \cdots, |N,-\rangle & \mbox{for} \ N \ \mbox{even}.
\end{array} \right.
\end{equation}
Using this basis and the driving term $\sigma_x f(t)$, the time-dependent Hamiltonian takes the form
\begin{equation}
\mathcal{H}(t) = \left( \begin{array}{ccccc}
E_0 & h_1 & 0 & \cdots & 0 \\
h_1 & E_1 & h_2 &   & \vdots \\
0 & h_2 & E_2 & \ddots & 0 \\
\vdots & &\ddots & \ddots & h_N\\
0 & \cdots & 0 & h_N & E_N 
\end{array} \right)
\end{equation}
where $\mathcal{H}_0 |v_n\rangle = E_n |v_n\rangle$ and
\begin{equation}
h_n(t) = \frac{1}{2} \left\{ \begin{array}{ll} \pm \sqrt{2} f(t) & \mbox{for} \ n=1, \\
\pm  f(t) & \mbox{for} \ n> 1. \end{array} \right.
\end{equation} 
The signs for $h_n(t)$ alternate depending on whether $N$ is odd $(-, +, -, \cdots,-)$ or $N$ is even $(+, -, +, \cdots,-)$. 

\subsection{Rotating Wave Hamiltonian}

By performing a transformation to an interaction picture with respect to $\mathcal{H}_0$, we find the rotating frame Hamiltonian
$\tilde{\mathcal{H}} = e^{i \mathcal{H}_0 t} \mathcal{H}(t) e^{-i \mathcal{H}_0 t} - \mathcal{H}_0$ is given by
\begin{equation}
\mathcal{\tilde{H}} = \left( \begin{array}{ccccc}
0 & h_1 e^{-i \omega_1 t} & 0 & \cdots & 0 \\
h_1 e^{i \omega_1 t} & 0 & h_2 e^{-i \omega_2 t} &   & \vdots \\
0 & h_2 e^{i \omega_2 t} & 0 & \ddots & 0 \\
\vdots &  &\ddots & \ddots & h_N e^{-i \omega_N t} \\
0 & \cdots & 0 & h_N e^{i \omega_N t} & 0
\end{array} \right)
\end{equation}
where $\omega_n = E_n - E_{n-1}$.

For the driving field
\begin{equation}
f(t) = \sum_{n=1}^N \Omega_n \cos (\omega_n t),
\end{equation}
the rotating wave approximation takes the form
\begin{equation}
h_n(t) e^{\pm i \omega_n t} \approx \frac{1}{4} \left\{ \begin{array}{ll} \sqrt{2} \Omega_1 & \mbox{for} \ n=1, \\
 \Omega_n & \mbox{for} \ n>1, \end{array} \right.
\end{equation}
where we have used $\cos(\omega t) = (e^{i \omega t} + e^{-i \omega t})/2$, neglected all terms with $e^{i (\omega_j - \omega_k) t}, \ j \ne k$ or $e^{i (\omega_j + \omega_k) t}$, and suppressed the alternating signs for the matrix elements.  The final rotating wave Hamiltonian is
\begin{equation}
\mathcal{H}_{\mbox{\scriptsize{RWA}}} = \frac{1}{4} \left( \begin{array}{ccccc}
0 &  \sqrt{2} \Omega_1 & 0 & \cdots & 0 \\
\sqrt{2} \Omega_1 & 0 & \Omega_2 &  & \vdots \\
0 & \Omega_2 & 0 & \ddots & 0 \\
\vdots & & \ddots & \ddots & \Omega_{N} \\
0 & \cdots & 0 & \Omega_{N} & 0
\end{array} \right)
\end{equation}

Setting 
\begin{equation}
\Omega_{n} = \Omega_0 \left\{ \begin{array}{cl}
\sqrt{2 N} & \mbox{for} \ n=1, \\
 2 \sqrt{n (N+1-n)} & \mbox{for} \ 1< n \le N,
\end{array} \right.
\end{equation}
the matrix elements reduce to
\begin{equation}
\langle n | \mathcal{H}_{\mbox{\scriptsize{RWA}}} | n+1 \rangle = \frac{\Omega_0}{2} \sqrt{n (N + 1 - n)}.
\end{equation}
Letting $j=N/2$ and $m = -j + n$, we have $\sqrt{n (N+1-n)} = \sqrt{(j+m) (j+1-m)}$, so that $\mathcal{H}_{\mbox{\scriptsize{RWA}}} = \Omega_0 (J_+ + J_-)/2 = \Omega_0 J_x$ where $J_x$ is an angular momentum operator for a particle of spin $j=N/2$.  

\subsection{Numerical Optimization}

The numerical optimization results reported in Fig. (2) in the text were obtained by a simplex search in Matlab ({\tt{fminsearch}}) over simulations of the time-dependent Schr{\"o}dinger equation to find the frequencies $\omega_n$ and Fourier amplitudes $a_{n}^{(k)}, b_{n}^{(k)}$ for the envelope functions $A_n(t)$ and $B_n(t)$.  Note that the transition frequencies $\omega_n/2\pi$ calculated from the eigenstates of $\mathcal{H}_0$ (with $\omega/2\pi = 6 \ \mbox{GHz}$ and $g/2\pi = 180 \ \mbox{MHz}$) are given by $\{ 6180, 5565.4, 6566.3, 5328.2 \} \ \mbox{MHz}$ .  The search results for several values of $T$ are reproduced in the following Tables:

\begin{table}[b]
\caption{Numerical optimization of the $N=4$ transition with $M=1$, with frequencies in MHz.} % title name of the table
\begin{center} 
\begin{tabular}{cccccc} 
\hline\hline
Time (ns) & $n$ & $\omega_n/2\pi$ & $a_n^{(1)}/2\pi$ & $b_n^{(1)}/2\pi$ & $1-\mathcal{F}$  \\ 
\hline
10 & 1 & 6252.7 & 209.2 & 61.0 & 0.22 \\
& 2 & 5947.2 & 87.5 & -169.2 & \\
& 3 & 6661.5 & 59.7 & -217.5 & \\
& 4 & 5450.1 & 286.4 & -159.1 & \\
\hline
20 & 1 & 6226.9 & 87.1 & 69.9 & 0.11 \\
& 2 & 5540.1 & 162.9 & -64.9 & \\
& 3 & 6572.4 & 99.9 & 26.1 & \\
& 4 & 5291.5 & 167.9 & -13.6 & \\
\hline
30 & 1 & 6200.4 & 19.1 & -69.6 & 0.031 \\
& 2 & 5557.3 & 56.9 & 29.8 & \\
& 3 & 6584.1 & 59.0 & -37.9 & \\
& 4 & 5299.7 & 32.7 & 101.4 & \\
\hline
40 & 1 & 6194.7 & 14.9 & -50.9 & 0.014 \\
& 2 & 5567.2 & 44.9 & -5.97 & \\
& 3 & 6561.8 & 46.0 & 27.5 & \\
& 4 & 5307.9 & 32.6 & 57.9 & \\
\hline
50 & 1 & 6195.3 & 7.54 & -52.3 & 0.005 \\
& 2 & 5579.1 & 20.7 & -67.1 & \\
& 3 & 6564.4 & 29.7 & -0.03 & \\
& 4 & 5330.2 & 4.83 & -70.3 & \\

\end{tabular}
\end{center}
\end{table}

\begin{table}[h] 
\caption{Numerical optimization of the $N=4$ transition with $M=3$, with frequencies in MHz.} % title name of the table
\begin{center} 
\begin{tabular}{cccccccccc} 
\hline\hline
Time (ns) & $n$ & $\omega_n/2\pi$ & $a_n^{(1)}/2\pi$ & $b_n^{(1)}/2\pi$ & $a_n^{(2)}/2\pi$ & $b_n^{(2)}/2\pi$ & $a_n^{(3)}/2\pi$ & $b_n^{(3)}/2\pi$ & $1-\mathcal{F}$  \\ 
\hline
10 & 1 & 6118.7 & 207.9 & 74.2 & -896.3 & -102.2 & -194.8 & -86.5 & $3.0 \times 10^{-8}$ \\
& 2 & 5679.1 & 340.1 & 111.3 & 240.4 & -369.4 & -6.03 & 91.9 &  \\
& 3 & 6694.1 & 365.9 & -202.2 & 55.36 & -282.3 & 41.7 & 100.7 & \\
& 4 & 5403.4 & 336.7 & 18.8 & -0.08 & -295.7 & 148.5 & 517.9 & \\
\hline
20 & 1 & 6248.7 & 134.4 & 101.0 & 76.8 & -28.2 & -36.6 & -5.58 & $2.9 \times 10^{-8}$ \\
& 2 & 5437.0 & 178.8 & 11.8 & -38.3 & 33.3 & 108.9 & 23.9 & \\
& 3 & 6666.4 & 121.1 & 0.43 & 84 & -47.1 & 73.9 & -27.4 & \\
& 4 & 5209.8 & 251.9 & 36.2 & -67.2 & 36.4 & 4.61 & -6.6 & \\
\hline
30 & 1 & 6193.9 & 11.0 & 14.3 & 39.4 & -22.9 & 7.92 & -54.8 & $2.3 \times 10^{-8}$ \\
& 2 & 5560.6 & 63.1 & 43.5 & -12.9 & -9.31 & 1.90 & 27.0 & \\
& 3 & 6591.5 & 82.2 & 7.62 & 11.7 & -3.17 & -21.2 & -16.5 & \\
& 4 & 5319.6 & 60.2 & 0.54 & 0.26 & -65.9 & 41.9 & -33.3 & \\
\hline
40 & 1 & 6162.1 & 4.66 & 13.2 & 28.3 & 29.9 & 0.06 & -0.13 & $5.3 \times 10^{-8}$ \\
& 2 & 5554.3 & 32.9 & 18.1 & 12.7 & -41.0 & -1.32 & -2.50 & \\
& 3 & 6550.0 & 48.8 & 34.4 & -31.5 & 18.7 & 5.39 & 38.6 & \\
& 4 & 5343.0 & 0.38 & -47.9 & 30.4 & -45.6 & 0.67 & -18.8 & \\
\hline
50 & 1 & 6182.8 & 5.33 & -22.6 & 0.22 & -7.19 & 23.6 & 20.1 & $7.0 \times 10^{-8}$ \\
& 2 & 5560.8 & 29.1 & 21.9 & 25.3 & 7.44 & -15.5 & 12.2 & \\
& 3 & 6574.8 & 23.5 & -18.3 & -9.12 & 4.15 & 18.0 & 8.40 & \\
& 4 & 5333.3 & 4.82 & -51.3 & 6.27 & -23.0 & 6.07 & -19.0 & \\

\end{tabular}
\end{center}
\end{table}

\end{widetext}

\end{document}